\documentclass[conference]{IEEEtran}
\IEEEoverridecommandlockouts
\usepackage{cite}
\usepackage{amsmath,amssymb,amsfonts}
\usepackage{algorithmic}
\usepackage{graphicx, caption, subcaption}
\usepackage{hyperref}
\usepackage{textcomp}
\usepackage{xcolor}
\def\BibTeX{{\rm B\kern-.05em{\sc i\kern-.025em b}\kern-.08em
    T\kern-.1667em\lower.7ex\hbox{E}\kern-.125emX}}
\begin{document}



\title{Investigating the Uncanny Valley Phenomenon Through the Temporal Dynamics of Neural Responses to Virtual Characters}

\author{\IEEEauthorblockN{Chiara Gorlini}
\IEEEauthorblockA{
\textit{Politecnico di Milano}\\
Milan, Italy \\
chiara.gorlini@mail.polimi.it}
\and
\IEEEauthorblockN{Laurits Dixen}
\IEEEauthorblockA{
\textit{IT-University Of Copenhagen}\\
Copenhagen, Denmark \\
ldix@itu.dk}
\and
\IEEEauthorblockN{Paolo Burelli}
\IEEEauthorblockA{
\textit{IT-University Of Copenhagen}\\
Copenhagen, Denmark \\
pabu@itu.dk}
}




\maketitle


\begin{abstract}
The Uncanny Valley phenomenon refers to the feeling of unease that arises when interacting with characters that appear almost, but not quite, human-like. First theorised by Masahiro Mori in 1970, it has since been widely observed in different contexts from humanoid robots to video games, in which it can result in players feeling uncomfortable or disconnected from the game, leading to a lack of immersion and potentially reducing the overall enjoyment.
The phenomenon has been observed and described mostly through behavioural studies based on self-reported scales of uncanny feeling: however, there is still no consensus on its cognitive and perceptual origins, which limits our understanding of its impact on player experience.
In this paper, we present a study aimed at identifying the mechanisms that trigger the uncanny response by collecting and analysing both self-reported feedback and EEG data.
\end{abstract}

\begin{IEEEkeywords}
Uncanny Valley, Virtual Characters, Video Games, EEG, ERP, Survey, Perception, Cognition
\end{IEEEkeywords}

\section{Introduction}

The concept of the Uncanny Valley (UV) refers to the phenomenon by which an artificial or virtual human-like entity is increasingly unsettling as it becomes more realistic but does not quite achieve a perfect human likeness. This effect can be observed in many media, including games, and has important implications for game design and player experience~\cite{tinwell_uncanny_2014}.

Studying this phenomenon can help game designers and developers to create more realistic and believable characters without crossing into the UV, which can lead to negative player reactions and reduced immersion. By understanding the factors that contribute to the  effect, teams can make informed decisions about how to design their characters and game worlds.

Moreover, by studying the  phenomenon, we can better understand the psychological processes that underlie our perception of virtual characters. Research has shown that the  effect is linked to our innate ability to recognize human facial expressions and interpret emotional cues~\cite{tinwell_facial_2011}. By studying the UV in games, we can gain insight into how our brains process visual information and how we form emotional connections with virtual characters.

The most common approach to studying the empirical basis of this phenomenon relies on self-reported subjective measurements, in which a series of stimuli is followed by questionnaires. These studies have shown repeatedly that some form of eeriness or unease appears to be triggered by different types of semi-realistic stimuli; however, the self-reported nature of these studies comes with some major limitations~\cite{wang_uncanny_2015}.

Self-reported feedback is subjective and can be influenced by a variety of factors, such as individual preferences, biases, and cultural differences. Additionally, self-reported feedback may not always accurately reflect a person's actual emotional response, as people may not always be aware of or able to articulate their feelings in a precise way. 

Moreover, the lack of uniformity of the stimuli and questionnaires used makes it difficult to compare results across studies. The questionnaires used to measure the UV effect do not always attempt to measure the same concept and define the uncanny feeling in very different ways. This can lead to inconsistencies in the results and make it challenging to draw definitive conclusions about the  effect.
Therefore, while self-reported studies can provide valuable insights into the UV effect, they should be supplemented with other objective measures, such as physiological responses or behavioural observations to obtain a more comprehensive understanding of how people react to artificial stimuli an understand its cognitive origins~\cite{cheetham_arousal_2015}.

In the last decade, in an attempt to address these limitations, researchers have started to investigate how brain activity is related to the UV. These studies have shown the potential to shed light on the causes of the effect based on more objective measurements such as EEG (electroencephalography) or fMRI (functional magnetic resonance imaging). Both are non-invasive methods to record, directly or indirectly, electrical activity in the brain allowing researchers to gain insight into how the brain processes and responds to stimuli~\cite{vaitonyte_scoping_2023}.

These studies have investigated both static and animated characters and have analysed their effect on the participants' brains primarily against different Event-Related Potentials (ERPs), with a few studies crossing brain activity measurements with behavioural variables~\cite{cheetham_human_2011,bagdasarian_eeg-based_2020}. 

Based on these results, in this paper, we present a two-steps study in which we first build a validated stimuli dataset based on self-reported subjective measurements collected through an online survey and then investigate the potential origins of these subjective measurements in a laboratory experiment in which we collect EEG responses to the validated stimuli.
By collecting both self-reported and psychophysiology measurements and by analysing the dynamics of the EEG response, we aim at finding objective evidence that can connect the two types of responses and that can either confirm or confute the different hypotheses on the origins of the UV phenomenon.

\section{Background}
Mashiro Mori theorised the Uncanny Valley phenomenon in 1970 to describe the unsettling feeling humans experience when looking at robots resembling human characteristics too closely ~\cite{mori_uncanny_2012}. The theory states that there is a non-linear relationship between the realism of an artificial character and its likability characterised by an evident dip at a level of realism that is almost but not quite perfect. While initially observed with humanoid robots, the rise of computer-generated images in video games and movies has opened a new area of application of the phenomenon, sparking new research aimed at finding behavioural and cognitive evidence of the UV~\cite{tinwell_facial_2011}. 


Over the years, different theories have formed to explain the UV phenomenon that can be roughly separated into two typologies. Firstly, the ones that explain the uncanny feeling as an instinctual response to stimuli early in brain perception (i.e. Pathogen Avoidance, Mortality Salience and Evolutionary Aesthetics)~\cite{wang_uncanny_2015}. The second class of theories explains the UV as a result of the cognitive process that occurs later in brain perception (i.e. Categorical Uncertainty, Violation of Expectation, Mind Perception, Dehumanization)~\cite{cheetham_arousal_2015, mustafa_how_2017, urgen_uncanny_2018}. These latter cognitive process hypotheses have received more credits in the literature, but there is still no clear evidence to fully support either of them. 

The primary mean of studying the UV is through self-reported subjective questionnaires. In these studies on the phenomenon, researchers manipulate a character’s level of human photo-realism and measure the character’s perceived humanness. Behavioural studies on UV mainly propose faces (still images and videos) as the stimuli to investigate. However, experiments differ in stimulus creation techniques for human-likeness modulations in robots, virtual agents, avatars, and perceptual scale modalities~\cite{diel_meta-analysis_2021}. 

Stimulus creation techniques are strictly controlled, ranging from artificial to entirely realistic characters (i.e. face distortion, realism render, and morphing)~\cite{macdorman_categorization-based_2017,lischetzke_topography_2017,sasaki_avoidance_2017}, or distinct entities based on the selection of existing computer-animated characters or robots~\cite{rosenthal-von_der_putten_how_2014}. 

Different perceptual exploratory constructs are used to describe the UV, such as eeriness, warmth, perceived threat, likability, and familiarity, whose relation can give insight into the phenomenon ~\cite{diel_meta-analysis_2021}. The dimensions are measured through a variety of terms used to define semantic differential scales. For example, MacDorman et al.~\cite{macdorman_reducing_2016} define the eeriness dimension as composed of 9 scales, each represented between two descriptors, Shin et al.~\cite{shin_uncanny_2019} consider the same dimension but use four descriptive scales, and Rosenthal–von der Pütten~\cite{rosenthal-von_der_putten_how_2014} do not consider eeriness but perceived threat as a main dimension. 

The variability between studies in terms of stimuli, scales and observations leads to the formulation of different results and causal explanations of the phenomenon. The UV shape varies between studies identifying linear relationships between human-likeness and affinity, 'weak uncanny valley', and 'strong uncanny valley' where the lowest affinity is predicted for high levels of human-likeness~\cite{katsyri_virtual_2019}. 
The different findings suggest that the UV is a multidimensional construct influenced by various factors, including stimulus creation techniques, non-trivial questionnaire terms, different levels of familiarity for characters, and other facets that might make entities unpleasant~\cite{diel_meta-analysis_2021}.

Although subjective questionnaires are a valuable tool for investigating the perception of characters, the complexity of the phenomenon makes them insufficient to explain the underlying causes of the UV. The human brain processes the faces between the onset of the visual stimulation and the behavioural response for several hundreds of milliseconds. Clarifying the exact time course of face processing by monitoring brain activity may give more insight into the reasoning behind the UV phenomenon.

There are few studies on brain activity related to UV in literature so far, see~\cite{vaitonyte_scoping_2023} for a recent comprehensive review. Electroencephalography (EEG) and functional magnetic resonance imaging (fMRI) have been used to find valuable biological markers and brain regions that might be involved in the face and object processing and the UV phenomenon.

Three EEG studies have been published using still images as stimuli. These studies use event-related potentials (LPP, N170, P200 and P300) as their main dependent variable for the different categories of stimuli~\cite{cheetham_arousal_2015, schindler_differential_2017}. ERPs are peaks in the electrical activity of the brain that reflect different stages of information processing and are usually identified by their polarity and time of occurrence.
Dynamic stimuli have also been used to test the hypothesised exacerbation of the UV curve. In EEG studies, the ERP adopted for such investigation is the N400~\cite{urgen_uncanny_2018}. 

One under-investigated aspect in the current literature is the comparison of early vs late processing~\cite{vaitonyte_scoping_2023, wang_uncanny_2015} and at what time the uncanny response occurs. This is an important question since the temporal dimension might help to discern the course of the feeling as cognitive mechanisms happen at different time points in the perceptual process.

In general, theories can be divided into perceptual processing theories (e.g. pathogen avoidance or mortality salience) and cognitive processing theories (e.g. violation of expectations or category uncertainty)~\cite{wang_uncanny_2015}. 
If the UV phenomenon relied on purely perceptual differences, we would expect to see the early ERP components being modulated when the subject is experiencing uncanniness instead of the later ones.

The present study focuses on two major ERP components: an early N170 component and a late N400 component. The N170 is specifically associated with processing human faces~\cite{rossion_n170_2012}. It is characterised by a low negative peak at the temporal lobe around 170 ms post-stimulus onset. The N170 is actually a special case of the visual ERP component called N1 but with a larger amplitude than non-face objects. Researchers are still discussing the neurological underpinnings of this 'N170 face effect', but there is a general agreement that the increased amplitude is evidence of more synchronised processing in the temporal lobe for facial stimuli~\cite{kappenman_erp_2011}.

In the context of UV, the N170 peak seems sensitive to facial human likeness and has a role in face categorisation. In particular, Schindler et al.~\cite{schindler_differential_2017} find a larger N170 amplitude for cartoon and natural human faces than for avatar faces. It is still uncertain how this early visual component is related to the UV phenomenon.
The other ERP component investigated in this study is the N400. This component has been widely used to identify semantic incongruities in the language -- e.g. ``I like my coffee with cream and sugar/dog''~\cite{kutas_thirty_2011}. The N400 is characterised by negative activity around the centro-parietal areas between 200 and 600 ms post-stimulation, particularly around 400 ms. 

It has been found in previous studies that increased N400 amplitude is related to robot-like movements of human-looking androids~\cite{urgen_uncanny_2018} and talking CG characters~\cite{mustafa_how_2017}. The common interpretation is that the artificial movements trigger an error correction response like a word in the wrong semantic context. It remains unknown if still images would elicit the same response, and be a general electrophysiological marker of the UV hypothesis or if it simply stems from the unexpectedness of artificial motions. 

The present study is inspired by the work of Wang et al.~\cite{wang_uncanny_2020}, in which they explored the reliability of the Dehumanisation Hypothesis through behavioural analysis. The Dehumanization hypothesis explains the uncanny feeling through the initial tendency to over-attribute a mind to nonhuman agents and then the perception of the anthropomorphise human replica as lacking humanness. 

Since their promising results demonstrate that the process of dehumanisation is more likely to account for the perceived uncanniness of android faces, the present study investigates the same hypothesis supported by an EEG study. By analysing the temporal dynamics of face processing through the study of both early and late ERPs, we aim at testing the Dehumanisation Hypothesis and uncovering the underlying mechanisms of the Uncanny Valley phenomenon.


\section{Methods}

The study is structured in two phases. In the first phase, we designed a questionnaire to collect self-reported responses to different kinds of virtual characters. This questionnaire aims at building a baseline of the different dimensions describing the "uncanny" feeling. We use the responses to fine-tune and validate the images that are  used for the second phase of the study.

In the study's second phase, we conduct a laboratory experiment, in which we expose a small sample of participants to a selection of previously validated virtual characters and analyse the dynamics of their EEG response in terms of event-related potentials (ERPs). By analyzing the ERPs of the participants in response to the virtual characters, we aim to identify which sequences of neural activations the different kinds of images trigger in the brain and, based on this, identify specific cognitive processes that are connected to the UV.

\subsection{Questionnaire}

\begin{figure*}
    \centering
    \begin{subfigure}[t]{.32\linewidth}
        \centering
        \includegraphics[width=\linewidth]{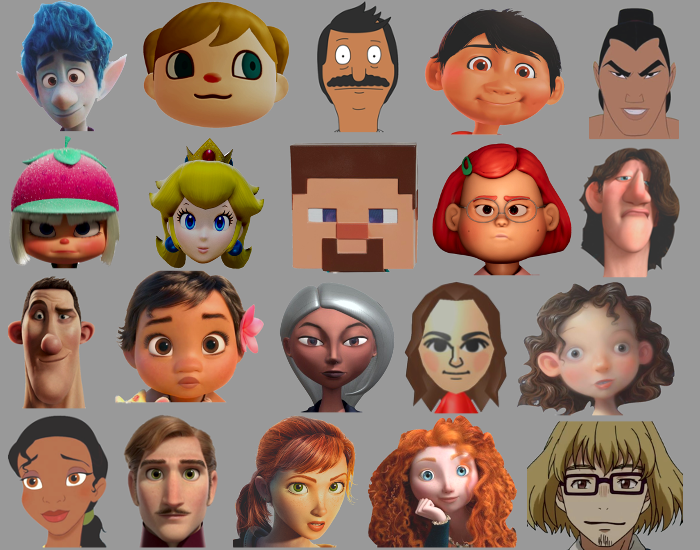}
        \caption{Unrealistic characters}
        \label{fig:dataset_animated}
    \end{subfigure}
    \begin{subfigure}[t]{.32\linewidth}
        \centering
        \includegraphics[width=\linewidth]{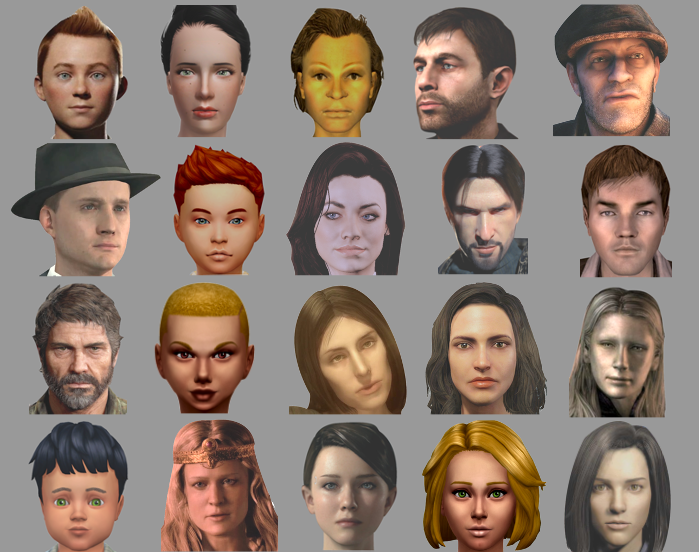}
        \caption{Semi-realistic characters}
        \label{fig:dataset_semireal}
    \end{subfigure}
    \begin{subfigure}[t]{.32\linewidth}
        \centering
        \includegraphics[width=\linewidth]{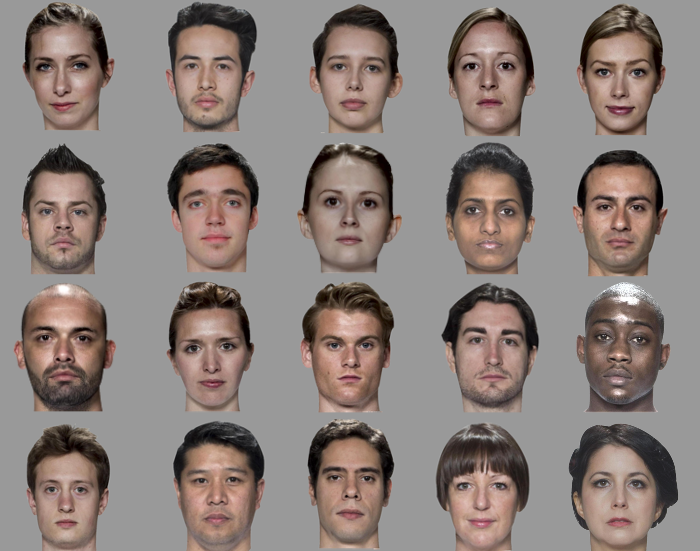}
        \caption{Realistic characters}
        \label{fig:dataset_human}
    \end{subfigure}
    \caption{Images used as stimuli in the EEG study. These images have been selected through a prior online questionnaire and they are organised into three categories: \textit{unrealistic}, \textit{semi-realistic} and \textit{realistic} characters.}
    \label{fig:dataset_faces}
\end{figure*}

The questionnaire\footnote{The source code of the questionnaire, the data and the statistical analysis are available at \url{https://github.com/itubrainlab/uncanny-face-questionnaire/tree/COG-2023}} is designed to collect reported feelings about a series of images. 
Every participant is first welcomed by a brief explanation of the study. After this first step, they are presented with a page that briefs them with some definitions and, after having given their consent, from this page, they are guided through a randomised sample of 20 images. For each image, their response is collected as a series of 7-point Likert scales. In the end, each participant is shown a debriefing page.

\subsubsection{Stimulus set}

Each participant is presented with a sequence of 20 images to rate, these images are a randomised sample from the full image dataset used in the questionnaire. The full dataset is composed of 109 images divided into three categories: \textit{unrealistic}, \textit{semi-realistic} and \textit{realistic}. The categorisation of the images is a direct consequence of the non-linear emotional response to realism that defines the UV phenomenon.

The first category (Figure~\ref{fig:dataset_animated}) includes unrealistic characters with exaggerated features often seen in the style of children's animated movies. These characters are rarely thought of as uncanny, despite clearly acting humanly. We picked the 36 characters belonging to this category from successful animation films and video games. The second category (Figure~\ref{fig:dataset_semireal}) includes animated synthetic characters with proportional human features designed to look realistic but don't achieve full realism. These are the characters which are more often described as uncanny, eliciting a negative emotional reaction regardless of the role they play. We picked 37 characters from video games and movies that have been described as uncanny on online communities and websites discussing the phenomenon -- e.g. \url{http://tvtropes.org}. The last category (Figure~\ref{fig:dataset_human}) includes a set of 36 face images of real people front-facing the camera. The images are selected from the Face Research Lab London Set~\cite{debruine_face_2017} with various gender and age combinations.
 
Inclusion criteria for the images are: high picture quality cropped and zoomed on the face of the character, and the character must show at least 3/4 of the face and have a neutral expression. The \textit{Unrealistic} and \textit{Semi-realistic} categories include characters from well-known digital productions, but rarely the main characters to avoid strong familiarity influencing the study. 

The images are processed to have the same resolution and size, their contrast and luminance are balanced, and each face is cut and pasted on a grey background. This is done to minimize low-level visual stimulus differences such as luminance and contrast to mitigate these differences between stimuli in the EEG part of the experiment. 

\subsubsection{Structure, questions and hypotheses}

According to the UV hypothesis~\cite{wang_uncanny_2015}, our three categories should fall on three different spots on the likability vs. human-likeness scale. The images in the \textit{Unrealistic} category should have low human likeness and medium to high likability. The human category should be high in both likability and human likeness. Lastly, the semi-realistic category should have high human likeness but low likability. This is the hypothesised outcome of the questionnaire. 

To operationalise the two dimensions of the UV effect, we construct three multi-item scales inspired by a recent literature review of Uncanny Valley~\cite{diel_meta-analysis_2021} studies. Each item consists of an anchor word at each end and a short description of the word.  The participants are asked to rate the character on a seven-point Likert scale with the two anchor words as the outer points for each item.

The structure of the scales and the items is the following:

\begin{itemize}
    \item \textbf{Realism} scale: \textit{Fictional/Real} [-3, +3], and \textit{Human-made/Human-like} [-3,+3];
    \item \textbf{Eeriness} scale: \textit{Ordinary/Eerie} [0,+6], \textit{Plain/Unsettling} [0,+6], \textit{Dull/Creepy} [0,+6], and \textit{Unemotional/Hair-raising} [0,+6];
    \item \textbf{Warmth} scale: \textit{Hostile/Friendly} [-3,+3], \textit{Grumpy/Cheerful} [-3,+3] and \textit{Cold-hearted/Warm-hearted} [-3,+3].
\end{itemize}

Seven of the nine items are meant to ascertain the ``uncanniness'' of the character; these items are divided into a \textit{Warmth} scale and an \textit{Eeriness} scale, largely inspired by previous work~\cite{ho_revisiting_2010,diel_meta-analysis_2021}. The \textit{Eeriness} scale is running from a neutral word to a negative word, intended to capture the negative aspect of uncanny feeling. The selection of terms aims at preserving the hypothesized construct of eeriness, which encompasses both fright and spine-tingling feelings. The specific terms are adapted from Ho and MacDorman~\cite{ho_revisiting_2010}, with a few adaptations to apply it to a static image dataset. 

The \textit{Warmth} scale is symmetrical around zero. Diel et al.~\cite{diel_meta-analysis_2021} find that this scale is not directly connected to the UV effect; however, we include it in the study as it has the potential to be an effective measure to discriminate the \textit{unreal} category~\cite{diel_meta-analysis_2021}. To measure the independent variable \textit{Realism}, we adapt the term to better suit animated characters specifically resulting in two items describing the realism of the character and how human-like it seems.

\subsubsection{Analysis}

Since the questionnaire is composed of three multi-item psychometric scales, the first step in the analysis consists in verifying the structure of these scales through confirmatory factor analysis~\cite{brown_confirmatory_2015}. First, to verify the existence of multi-item factors, we conduct a Bartlett’s test of Sphericity; second, we calculate the Kaiser-Meyer-Olkin (KMO) Measure of Sampling Adequacy to examine the strength of the partial correlations between the items. Finally, we fit a Confirmatory Factor Analysis model to estimate the loadings of the items with respect to the scales. 

For each scale verified in the aforementioned analysis, we calculate their means across categories and verify the presence of significant differences through a series of non-parametric Mann-Whitney U tests. Based on the results of this analysis, we select a subset of 20 images for each category that perform better in terms of "uncanny" feeling to be used for the EEG Experiment. The reason for this sampling is to make sure that only images that are validated to follow the UV curve are used as stimuli in the experiment; this way, we can increase the likelihood to capture the phenomenon in the participants' psychophysiological responses. 

\subsection{EEG Experiment}

The purpose of this experiment is to collect the EEG response to the images selected through the survey. The testing procedure\footnote{The source code of the EEG experiment and the analysis are available at \url{https://github.com/itubrainlab/uncanny-face-eeg/tree/COG-2023}} to collect this data is the following: first, the participants are welcomed into the testing room and received a short briefing; second, they are seated in front of a screen and prepared for the recording. After the preparation, each participant goes through two 20 minutes of recording sessions separated by a short break to prevent fatigue.
    
Each session contains 240 trials in randomised order. With 20 images in each stimulus condition, each participant sees every image four times. The participants are instructed to restrain from blinking during the stimulus time, and breaks of 10 seconds every 20 trials are added to minimise eye strain. 

During each trial, the participant is first exposed to a centered fixation cross for 500 ms followed by a period of stimulus onset asynchrony (SOA) randomised between 300 and 600 ms. Afterwards, the stimulus is shown for 700 ms. The participant is then asked to rate the character just seen in terms of perceived animacy on a 1-3 scale. The participant uses their left hand and has a maximum of 2000 ms to rate. 

The purpose of the task is to keep the participant alert; we chose 'animacy' as it relates directly to the dehumanisation hypothesis~\cite{wang_uncanny_2015, wang_uncanny_2020}, this way the participant hopefully was constantly thinking about the animacy of the character, which should drive the dehumanisation process. Furthermore, the task was chosen to be as simple as possible to not disturb the EEG signal with motor or pre-motor activity in preparation for the reporting movement. The analysis of this behavioural task is not of major importance in this article's analysis, since we have poor granularity with only three options. 

\subsubsection{Equipment}
\begin{figure}[t]
        \centering
        \includegraphics[width=0.6\linewidth]{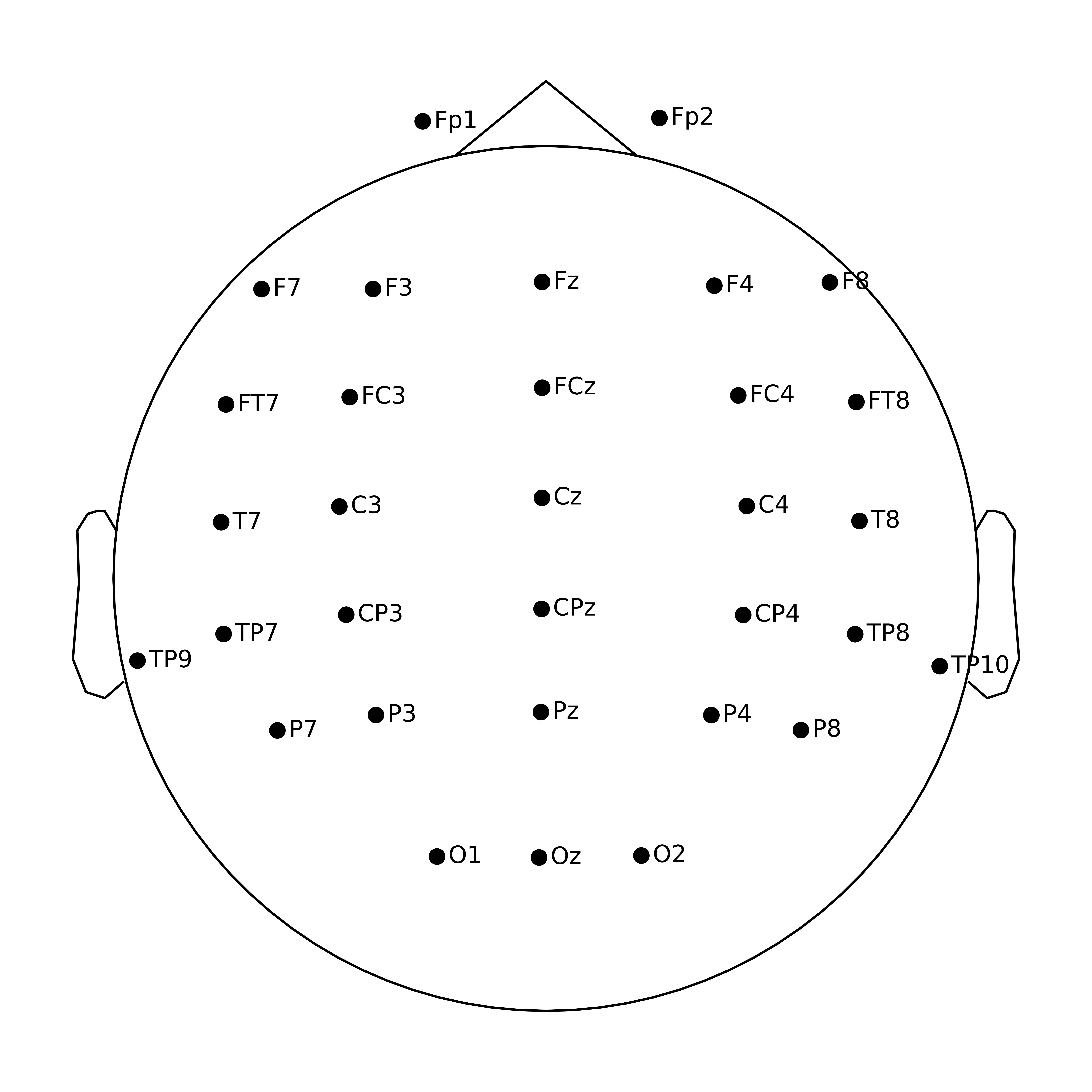}
        \caption{EEG montage displaying the placement of the 32 electrodes.}
        \label{fig:montage}
\end{figure}

The EEG recording was executed using a g.tec\footnote{https://www.gtec.at/product/gusbamp-research/} 32 channel g.USBAMP amplifier, with g.LADYbird active electrodes arranged in accordance with the 10/20 system~\cite{jasper_ten-twenty_1958} and referenced to an earlobe electrode. The placement of the electrodes can be seen in Figure~\ref{fig:montage}. Data was recorded using MathWorks Simulink\footnote{https://www.mathworks.com/products/simulink.html} with a $256$ Hz sampling frequency filtered with a $0.1$ Hz high-pass filter and a $50$ Hz Notch filter to remove drift and power line noise. 

\subsubsection{Pre-processing}

After the recording, the EEG data is further filtered using a FIR band-pass filter at 1 Hz to 35 Hz. Then flat and noisy channels are detected and removed from further analysis with the RANSAC algorithm, following the procedure of the PREP EEG pre-processing pipeline~\cite{bigdely-shamlo_prep_2015}. The eye blink artefacts are removed through Indepependent Compoment Analysis (ICA)~\cite{tandle_classification_2016}, training the ICA on four-second intervals~\cite{delorme_independent_2012}. 
The data was split into epochs from $-200$ ms before to $700$ ms past the stimulus onset and baselined on the pre-stimulus part of the epoch. Again following the PREP pipeline, the Autoreject~\cite{jas_autoreject_2017} algorithm is used to detect and interpolate or reject bad epochs. EEG pre-processing and analysis are based on the MNE python package~\cite{gramfort_meg_2013}. 

\subsubsection{Small N studies and within-participant design}

The EEG analysis in this study features a relatively small number of participants, which would normally limit the generalisability of the conclusions drawn from the data. However, relying on within-participant analysis rather than group-level analysis can increase the robustness of the conclusions significantly~\cite{smith_small_2018}. Using the singular participant as the replicational unit decreases the risk of type I errors since we rely on the same pattern showing up in each participant independently. 

\subsubsection{ERP analysis}\label{sec:methods_erp}


 Two ERP components analysed in this study are the N170 and N400. An upside of small N studies is that they simplify the customisation of the analysis to individual differences between participants; in this study, for instance, we can personalise the time windows of analysis for the N170 components of each participant to accommodate the individual variability of the onset and peak of the component. 
 To prevent leakage into our statistical analysis, the time windows are picked based on the evoked potential averaged on all conditions. 
 
 Contrarily to the N170, the N400 component is expected to be present only in one condition; therefore, we cannot pick time windows based on the average signal and we cannot have individualised time windows. Instead, the time window was set at 450-600 ms for all participants. To measure the N170, we use the singular electrode P8, which has been shown to consistently show the N170 face effect, re-referencing the signal to the average amplitude~\cite{rossion_n170_2012}. To measure the N400, we use the centro-parietal electrodes: CPz, CP3, CPz, CP4, P3, Pz, P4~\cite{kutas_thirty_2011}.

 We are using the \textit{realistic} human faces as the baseline condition, so the hypotheses are stated as a difference to this condition. We are testing two main hypotheses: 
 \begin{enumerate}
     \item[H1] There is a significantly stronger N170 response to \textit{unrealistic} characters but not the \textit{semi-realistic} ones when compared to \textit{realistic} characters.
     \item[H2] There is a significantly stronger N400 response to \textit{semi-realistic} characters but not the \textit{unrealistic} ones when compared to \textit{realistic} characters.
 \end{enumerate}


\section{Results}

Following the structure of the study, this section is divided into two parts. In the first one, we present the results collected and analysed from the questionnaire and we describe the dataset selected for the EEG experiment. In the second part, we present the result and the analysis of the EEG data collected in the experiment.

 \begin{table}[t]
    \centering
\begin{tabular}{rrrr}
\hline
 & Realism & Uncanny & Warmth \\
 \hline
 Fictional/Real & 2.60 & 0.00 & 0.00 \\
 Human-made/Human-like & 2.06 & 0.00 & 0.00 \\
 Eerie/Ordinary & 0.00 & 1.56 & 0.00 \\
 Unsettling/Plain & 0.00 & 1.68 & 0.00 \\
 Creepy/Dull & 0.00 & 1.57 & 0.00 \\
 Hair-raising/Unemotional & 0.00 & 1.21 & 0.00 \\
 Hostile/Friendly & 0.00 & 0.00 & 1.37 \\
 Grumpy/Cheerful & 0.00 & 0.00 & 1.21 \\
 Cold-hearted/Warm-hearted & 0.00 & 0.00 & 1.31 \\
\hline
\end{tabular}
    \caption{Confirmatory factor analysis loadings of the scales and their items in the questionnaire.}
    \label{tab:cfa_loadings}
\end{table}


Through the online questionnaire, we collected responses from 273 participants. The participants' age is between 17 and 57 years with a mean of approximately 29 with around 56\% of the sample being composed of non-native English speakers. Each image has been rated between 20 and 50 times.

In the confirmatory factor analysis, the Bartlett’s test of Sphericity results in a chi-square value of $18964.41$ p-value equal to $0.0$. This indicates that the dataset has a lower dimensionality compared to the number of items, a necessary condition of the existence of the multi-item scales. The result of the Kaiser-Meyer-Olkin (KMO) measure of sampling adequacy is $0.815$, which indicates the presence of strong partial correlations between the items. Finally, the confirmatory factor analysis shown in Table~\ref{tab:cfa_loadings}, confirms the presence of three independent factors: \textit{Realism}, \textit{Eeriness} and \textit{Warmth}, corresponding to the three multi-items scales designed in the questionnaire. For each factor, we calculate a score equal to the sum of the items' values, with a range between $-6$ and $+6$ for the \textit{Realism} scale, $0$ and $24$ for the \textit{Eeriness} scale, and $-9$ and $+9$ for the \textit{Warmth} scale. 

Based on the Eeriness scale, a subset of 20 images from each category is selected. This subset contains images that maximise the difference in reported eeriness feeling between the \textit{semi-realistic} category and the other two. 
Through this criterion, we aim at selecting images with a stronger potential to generate an uncanny response that follows the canonical Uncanny Valley. 

A non-parametric Mann-Whitney U test was conducted on these images to verify whether the categories reflected their respective labels. Table\ref{tab:subjresults} shows the results obtained. 

\begin{table}[h]
    \centering
\begin{tabular}{rrrr}
\hline
 & p-value & Median Score & Score Scale\\
 \hline
 \textit{Eeriness} & $<0.01$ & 7, 12, 3 & [0,+24] \\
 \textit{Warmth} & $<0.01$ & 4, -1, 1 & [-9,+9] \\
 \textit{Realism} & $<0.01$ & -6, -4, 6 & [-6,+6] \\
\hline
\end{tabular}
    \caption{\textit{Non-parametric Mann-Whitney U test results.} The Score Scale is computed by summing the items' values in the specific scale. The Median Scores are reported for \textit{Unrealistic}, \textit{Semi-realistic} and \textit{Realistic} categories, respectively.}
    \label{tab:subjresults}
\end{table}

\subsection{ERP analysis}

\begin{figure}[t]
    \centering
    \includegraphics[width=\linewidth]{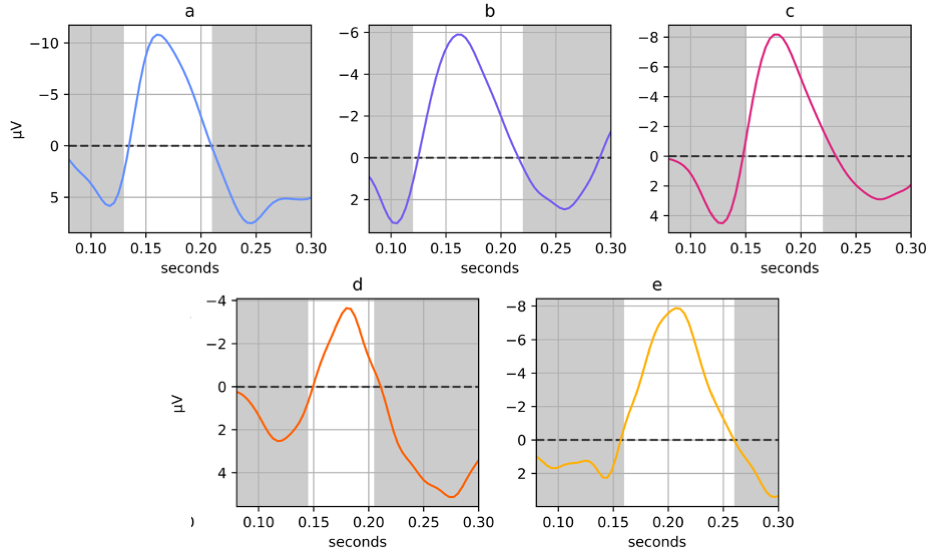}
    \caption{N170 time windows estimated for the five participants in the EEG experiment.}
    \label{fig:170window}
\end{figure}

Five people (1 female) participated in the EEG experiment. They were all sampled from the staff at the IT University of Copenhagen. Crucially, they were blind to the experiment's purpose and hypothesis and had also not participated in the questionnaire described above. The participants are labelled '\textit{a}', '\textit{b}', '\textit{c}', '\textit{d}' and '\textit{e}' in the remainder of the text\footnote{The dataset, containing the collected data and the stimuli, is available at \url{https://doi.org/10.5281/zenodo.7948158}}.

For each participant, we analyse the characteristics of their cognitive response based on the intensity of their N170 and N400 ERPs. As described in Section~\ref{sec:methods_erp}, the time windows for the N170 component are personalised for each participant based on their average ERP across all experimental conditions. The selected time windows can be seen in Figure~\ref{fig:170window}.
The analysis of the two ERPs is performed within-participant and replicated across participants. The statistical tests are performed over the trials that each participant goes through, divided into the three image categories, giving 160 trials per category and participant.

    \begin{figure}[t]
        \centering
        \includegraphics[width=0.9\linewidth]{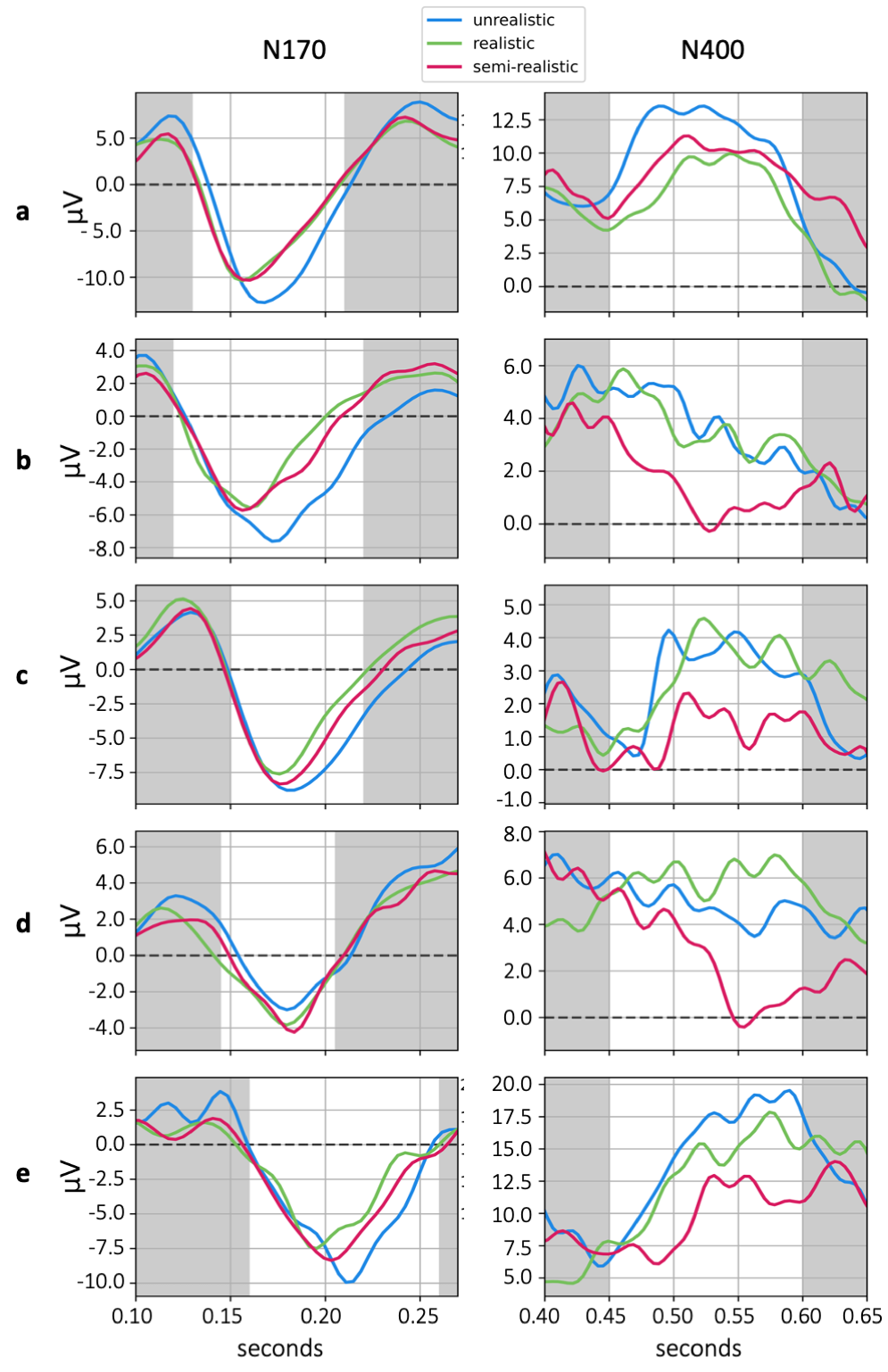}
        \caption{ERP waves in the N170 and N400 time windows for the three categories. Each row corresponds to one participant.}
        \label{fig:evokeds}
    \end{figure}

The first hypothesis is that the N170 amplitude will be stronger (more negative, since the component has a negative pole) in the animated condition, but not in the semi-realistic condition when compared to the human condition.
From a visual inspection of the ERP curves in Figure~\ref{fig:evokeds}, it is immediately apparent that there is a difference between the categories in the N170.
This is also confirmed through an independent sample t-test, as characters belonging to the \textit{unrealistic category} elicit a significantly ($p<0.05$) stronger N170 than characters in the \textit{realistic} category in 4 out of 5 participants (all but participant '\textit{d}'). 

The second hypothesis is that \textit{semi-realistic} characters will be the only ones eliciting an N400 response representing as their later cognitive processing contradicts the earlier human face processing. This difference is apparent in the ERP curves in Figure~\ref{fig:evokeds} and an independent sample t-test shows a statistically significant difference in the intensity of the N400 component ($p<0.05$) for the \textit{semi-realistic} compared to the \textit{realistic} characters in 4 out of 5 participants (all but participant '\textit{a}'). We do not find significant evidence for an N400 response instead when analysing the response to \textit{unrealistic} characters. 



\section{Discussion}


The results of the ERP analysis show that, in 4 out of 5 participants, the \textit{unrealistic} characters elicit a stronger N170 face effect when compared to the \textit{realistic} ones. In contrast, only one participant demonstrates the same increased N170 effect in the \textit{semi-realistic} condition. This suggests that at this stage of processing facial stimuli, the \textit{unrealistic} faces are processed differently than the \textit{realistic} and \textit{semi-realistic} faces. The N170 is considered to mark the brink of conscious processing of facial stimuli and not the immediate recognition of a face~\cite{rossion_understanding_2014}. At this stage of processing, the participant is discerning the high-level characteristics of the face, such as sex, age, and expression, based on individual features. 
It seems, therefore, that the unusual composition of the unrealistic characters engages the face-specific brain processing more, as it might be harder to discern high-level characteristics of the character. However, this also suggests that the UV phenomenon does not stem solely from this step in facial processing, since the unrealistic and semi-realistic characters chosen from the survey were specifically chosen to elicit the lowest and highest uncanniness response, respectively.

By looking at the later event-related potential response, we can see that 4 out of 5 participants have a significantly stronger N400 response to \textit{semi-realistic} characters when compared to the \textit{realistic} one and no participant had a corresponding response to the \textit{unrealistic} characters. 
This suggests that some form of semantic error correction is happening when looking at these almost-but-not-quite-human faces. As suggested by the N170 analysis, the early processing of semi-realistic faces resembles the processing of human faces due to the similar proportions. But as the participant realizes the face is not actually a human face, an error-correcting response is necessary, as the character is indeed not an actual human. This interpretation is in line with the prediction of the dehumanization hypothesis˜\cite{wang_uncanny_2015} as an expectation of animacy and therefore humanity is generated in the early stage of visual processing, which triggers an error correction afterwards, as one realizes the character lacks actual inner life. We suggest the N400 response marks the beginning of the error-correcting mechanism of this dehumanizing process.

In both ERP analyses, one of the five participants did not follow the pattern of the other four. In group-level analyses with a higher number of participants, these incongruities would go unnoticed if the effect was strong enough. However, in a small N study like ours, where each participant essentially functions as a replication, deviant results warrant a discussion. In the N170 analysis, the participant '\textit{d}' is the outlier. Looking closer at the participant's N170 (Figure~\ref{fig:170window} and \ref{fig:evokeds}), one can see the amplitude of the N170 has a lower amplitude and a shorter time-window compared to other participants. This observation might be the consequence of the single-electrode nature of this measurement or might be the manifestation of an inter-participant difference. We believe further studies would be necessary to verify the origin of this ERP outlier.

In the N400 analysis another participant, participant '\textit{a}', does not show a significantly negative peak in the later time window. Based on the rating of animacy collected from each participant, we can say that this person rarely gave a medium score on animacy regardless of realism, particularly in the \textit{semi-realistic} condition, when compared to other participants. The lack of an N400 effect could stem from this participant simply not perceiving the \textit{semi-realistic} characters as resembling humans. A solid link between behavioural and neurological results cannot be established from this single participant alone, but it provides a possible reconciliation of the result's deviancy from the rest of the group.
To draw a clearer picture of the individual differences in terms of UV response, future studies in this direction should capture individual behavioural responses contextually to the EEG measurements. A larger-scale study including this kind of data has the potential to lead to a better understanding of the link between subjective reports and EEG signals.

On a final note, this experiment follows a well-established protocol to capture event-related potential neural responses, which requires long sessions with a large number of stimuli repetitions to achieve the best signal-to-noise ratio. However, the elevated robustness and validity of the measurements given by this design come at the expense of a potentially lower ecological validity, as there is a clear gap between the experimental condition and the real context in which a player would interact with digital characters. Bridging this gap, with more realistic experimental protocols, could unlock a more widespread usage of neurophysiological data in player experience modelling, leading potentially to more accurate and empirically-grounded models.

\section{Conclusions}

In this study, we investigate the cognitive origins of the Uncanny valley phenomenon by collecting behavioural and EEG responses to a set of images in a survey and a laboratory experiment. Through the survey, we select a set of images that elicit a behavioural response compatible with the Uncanny Valley, these images are used in the EEG experiment to capture the neural response of the participants at different locations and time frames.
The data and the analysis give a strong indication that the phenomenon is a consequence of a mismatch in the early and later cognitive processing, in which an \textit{uncanny} face is first detected as human and later ``corrected''. While a number of open questions still remain to be answered, especially in terms of individual differences between the participants and how to study them, we believe the results contribute to a better understanding of the Uncanny Valley phenomenon and showcase the applicability of neurophysiological studies to player experience analysis.

\bibliographystyle{plain}
\bibliography{references}

\end{document}